\documentclass[a4paper,twocolumn,preprintnumbers,citeautoscript,newabstract,aps,nofootinbib]{revtex4-2} 
\usepackage{amsmath,amsfonts,amssymb,graphics}
\usepackage[normalem]{ulem}
\usepackage{graphicx}
\usepackage{units}
\usepackage{mathrsfs}
\usepackage{bbm}
\usepackage{pifont}
\usepackage{braket}
\usepackage{units}
\usepackage[dvipsnames]{xcolor}
\usepackage{mathtools}
\usepackage{xspace}
\usepackage{booktabs}

\usepackage[bookmarks=false,hyperfootnotes=false]{hyperref}

\definecolor{nicered}{rgb}{0.5,.0,.0}
\definecolor{darkblue}{rgb}{0,.1,.9}
\definecolor{lightblue}{rgb}{0,.1,.6}
\definecolor{darkgreen}{rgb}{0.0,0.2,0.0}
\hypersetup{colorlinks, citecolor=darkgreen ,linkcolor=darkblue, urlcolor=lightblue}

\newcommand{\SO}[1]{\ensuremath{\mathrm{SO}(#1)}}
\newcommand{\SU}[1]{\ensuremath{\mathrm{SU}(#1)}}

\newcommand{\U}[1]{\ensuremath{\mathrm{U}(#1)}}

\newcommand{\rep}[2][]{\ensuremath{\boldsymbol{#2}#1}}

\renewcommand{\bar}[1]{\overline{#1}}

\newcommand\varpm{\mathbin{\vcenter{\hbox{%
  \oalign{\hfil$\scriptstyle+$\hfil\cr
          \noalign{\kern-.3ex}
          $\scriptscriptstyle({-})$\cr}%
}}}}
\newcommand\varmp{\mathbin{\vcenter{\hbox{%
  \oalign{$\scriptstyle({+})$\cr
          \noalign{\kern-.3ex}
          \hfil$\scriptscriptstyle-$\hfil\cr}%
}}}}

\definecolor{darkgreen}{HTML}{109930}

\begin{document}

\title{\textbf{\boldmath\Large\mbox{Electroweak hierarchy from conformal and custodial symmetry}\unboldmath}}

\author{Thede de Boer}\email[]{thede.deboer@mpi-hd.mpg.de}
\author{Manfred Lindner}\email[]{lindner@mpi-hd.mpg.de}
\author{Andreas Trautner}\email[]{trautner@mpi-hd.mpg.de}

\affiliation{\vspace{0.2cm}Max-Planck-Institut f\"ur Kernphysik, Saupfercheckweg 1, 69117 Heidelberg, Germany}

\begin{abstract}
We present  ``Custodial Naturalness'' as a new mechanism to explain the separation between the electroweak (EW) scale and the scale of potential ultraviolet completions of the Standard Model~(SM). We assume classical scale invariance as well as an extension of the SM scalar sector custodial symmetry to~\SO{6}. This requires a single new complex scalar field charged under a new $U(1)_\mathrm{X}$ gauge symmetry which partially overlaps with~$B-L$. Classical scale invariance and the high-scale scalar sector \SO{6} custodial symmetry are radiatively broken by quantum effects that generate a new intermediate scale by dimensional transmutation. The little hierarchy problem is solved because the Higgs boson arises as an elementary (i.e.~non-composite) pseudo-Nambu-Goldstone boson (pNGB) of the spontaneously broken $\SO{6}$ custodial symmetry. The minimal setting has the same number of parameters as the SM and predicts new physics in the form of a heavy $Z'$ with fixed couplings to the SM and a mass of $m_{Z'}\approx4-100\,\mathrm{TeV}$, as well as a light but close-to invisible dilaton with a mass $m_{h_\Phi}\approx75\,\mathrm{GeV}$. 
\end{abstract}

\maketitle
\widowpenalty100000\clubpenalty100000
\section{Introduction}\enlargethispage{1cm}
The SM exhibits classical scale symmetry, explicitly broken only by the EW scale Higgs mass. While quantum corrections could spontaneously generate the EW scale via dimensional transmutation \'a la Coleman-Weinberg~(CW)~\cite{Coleman:1973jx},
the simplest incarnation of this mechanism is excluded as it requires a small top Yukawa coupling and a Higgs mass of the order~$10\,\mathrm{GeV}$~\cite{Weinberg:1976pe,Gildener:1976ih}. Instead, dimensional transmutation could happen in an extended scalar sector and indirectly induce the EW scale~\cite{Hempfling:1996ht, Meissner:2006zh,Espinosa:2007qk,Chang:2007ki,Foot:2007as,Iso:2009ss} which, however, typically introduces a little hierarchy problem through the Higgs portal.
We show that this can be avoided if the generation of the new scale also spontaneously breaks an extended custodial symmetry. If the Higgs is a pNGB of the spontaneously broken custodial symmetry, the transmission of the new scale to the SM is naturally suppressed.

Having the Higgs as a pNGB of an approximate global symmetry is a typical feature of strong coupling solutions to the hierarchy problem such as composite Higgs~\cite{Kaplan:1983fs,Kaplan:1983sm,Georgi:1984af,Dugan:1984hq}, little Higgs~\cite{Arkani-Hamed:2001nha,Arkani-Hamed:2002sdy,Arkani-Hamed:2002ikv} or twin Higgs models~\cite{Chacko:2005pe,Barbieri:2005ri,Chacko:2005vw}. However, even in minimal scenarios~\cite{Agashe:2004rs, Cacciapaglia:2014uja}, 
and despite taking into account enhanced custodial and/or conformal symmetries~\cite{Katz:2005au,Galloway:2010bp,Ahmed:2023qsm},
generating the top Yukawa coupling requires introduction of top partners or fine tuning.

In contrast to all of these, our model only contains elementary fields in a weak coupling regime and a marginal top coupling like~\cite{Alanne:2014kea,Gertov:2015xma}, compared to which, however we extend the gauge group, use a simpler scalar sector and incorporate classical scale invariance similar to~\cite{Iso:2009ss,Iso:2009nw,Oda:2015gna,Das:2015nwk,Das:2016zue}. As a real novelty, we highlight the importance of high-scale custodial symmetry.

Our minimal model couples to the SM via Higgs and gauge kinetic mixing portal, while the neutrino portal can be populated in extensions. We will see that it is the interplay of portals that makes our scenario of spontaneous scale generation particularly worthwhile.

\section{General Idea}
We amend the SM by a complex scalar field $\Phi$ charged under a new $\U{1}_\mathrm{X}$ gauge symmetry under which also the SM Higgs $H$ is charged. At the ultraviolet cutoff scale of the model, which we generically take to be $M_{\mathrm{Pl}}$, the scalar potential is assumed to be both, classically scale invariant as well as symmetric under a $\SO{6}$ custodial symmetry,\footnote{%
We refer to ``custodial symmetry'' as the full symmetry of the scalar potential, broken by gauge and Yukawa interactions. This aligns with the common statement of ``$\SO{4}$'' for the SM, not only its remainder after symmetry breaking~\cite{Sikivie:1980hm}. We remark that the actual groups in both cases can contain disconnected components and/or multi covers of the $\SO{n}$ algebra.
Additional quartic terms in the scalar potential allowed for a generic $\rep{6}$ of $\SO{6}$ are prohibited by gauge invariance.
} 

\begin{equation}\label{eq:SymV}
	\Big. V(H,\Phi)~=~\lambda\left(|H|^2+|\Phi|^2\right)^2\;~\text{at}~\mu=M_\mathrm{Pl}\,.
\end{equation}
Both, the conformal as well as the custodial symmetry of the potential are broken by quantum effects and this determines the phenomenology at scales below the \mbox{UV~cutoff}. The breaking of conformal symmetry via the trace anomaly gives rise to dimensional transmutation via the CW mechanism. This induces quadratic terms and VEVs for~$H$ and~$\Phi$ that are naturally exponentially suppressed compared to the cutoff. 

\begin{figure}[t]
	\centering
	\includegraphics[width=0.5\textwidth]{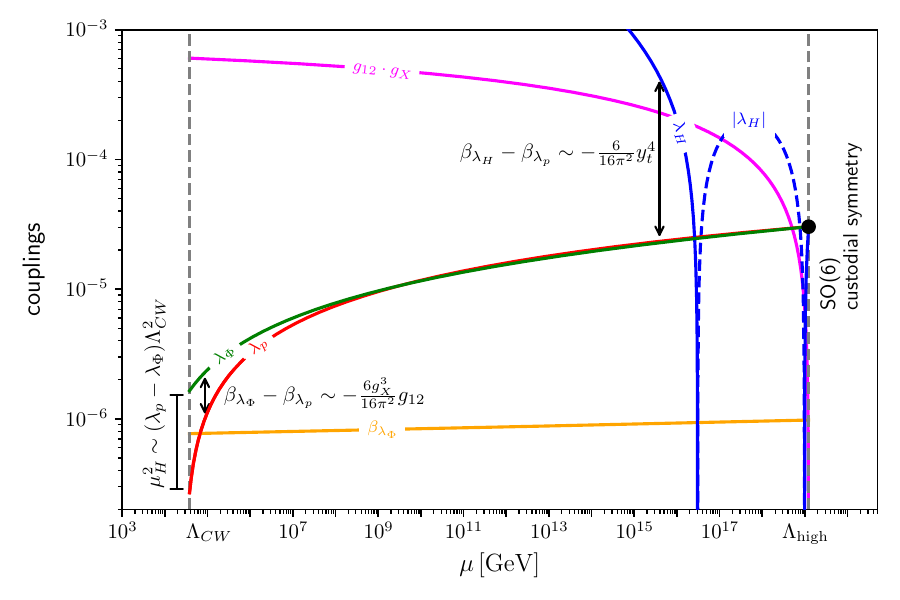}
	\caption{\label{fig:running}
	Running of scalar quartic couplings below a high scale $\Lambda_\text{high}$ where $\lambda=\lambda_H=\lambda_p=\lambda_\Phi$ with $\SO{6}$ custodial symmetry. Typically, $\lambda_p\lesssim 10^{-3}$ even at the high scale. $\lambda_H$~can cross zero and we show $|\lambda_H|$ as dashed line.
 The approximate custodial symmetry protects the difference of $\lambda_\Phi-\lambda_p$ and thereby the suppression of the EW scale.}
\end{figure}
To understand the role of physical fields, their masses and couplings, 
it is instructive to think about the case with unbroken custodial symmetry.
If custodial symmetry were to be exact, it would be spontaneously broken like $\SO{6}\rightarrow\SO{5}$ once the $H-\Phi$ system gets a VEV from spontaneous scale symmetry breaking. This would leave us with a massive dilaton (radial mode and pNGB of spontaneous scale symmetry breaking) and $5$ massless Nambu-Goldstone modes. 
In the realistic setup, $4$ of the scalars are, in fact, would-be Goldstone bosons that are eaten by the gauge bosons of spontaneously broken $\SU{2}_{\mathrm{L}}\times \U{1}_{\mathrm{Y}}$ and $\U{1}_\mathrm{X}$. 
The remaining scalars are the dilaton, as well as a pNGB associated with the spontaneous breaking of $\SO{6}$ custodial symmetry which closely resembles the SM Higgs field.

The $\SO{6}$ custodial symmetry is explicitly broken by SM gauge interactions and (analogously to $\U{1}_\mathrm{Y}$ in the SM) by the new $\U{1}_\mathrm{X}$ gauge coupling $g_X$, as well as most dominantly by the top Yukawa coupling $y_t$. 
We write the conformal scalar potential as 
\begin{equation}
	V_\text{tree}(H,\Phi)~=~\lambda_H |H|^4 +2\,\lambda_p|\Phi|^2 |H|^2+ \lambda_\Phi|\Phi|^4\;.
\end{equation}	
The dominance in custodial breaking of $y_t$ drives $\lambda_H$ to a large value while $\lambda_p$ and $\lambda_\Phi$ stay small and close to each other, see Fig.~\ref{fig:running}. This results in a flat direction of the potential that points predominantly along the $\Phi$ field direction, thereby ensuring a hierarchy of VEVs $\langle H\rangle\ll\langle\Phi\rangle$.
This establishes $\langle\Phi\rangle=:\nicefrac{v_\Phi}{\sqrt{2}}$ as the intermediate scale of spontaneous scale and custodial symmetry violation, while the EW breaking $\langle H\rangle=:\nicefrac{v_H}{\sqrt{2}}$ is suppressed.

Crucially, there needs to be a new, yet subdominant, source of explicit custodial symmetry violation which splits \mbox{$\lambda_\Phi-\lambda_p$} to the correct sign and size in order to obtain a realistic EW scale.
In the minimal case, this is 
the \mbox{$\U{1}_\mathrm{X}-\U{1}_\mathrm{Y}$} gauge kinetic mixing $g_{12}$.\footnote{Diagonalizing the kinetic term shifts the gauge kinetic mixing~\cite{Galison:1983pa,Holdom:1985ag} from $\epsilon\,F^{\mu\nu}F'_{\mu\nu}$ into a triangular gauge coupling matrix whose off-diagonal entry is given by $\frac{\epsilon g_Y}{\sqrt{1-\epsilon^2}}=:g_{12}$, see e.g.~\cite{Pan:2018dmu}.}
Alternative sources of custodial symmetry breaking (e.g.\ new Yukawa couplings) are possible in extensions of our minimal scenario, as will be discussed in detail in~\cite{DeBoer:2024xx}.

\enlargethispage{0.4cm}
The masses of the physical real scalars hosted in $\Phi$ and $H$ are approximately given by
\begin{subequations}\label{eq:mhPhi}
\begin{align}
m^2_{h_\Phi}&~\approx~\frac{3\,g_X^4}{8\pi^2}\,v^2_{\Phi}\;,\quad\text{and}& \\
m^2_h&~\approx~2\left[\lambda_\Phi \left(1+\frac{g_{12}}{2\,g_X}\right)^2-\lambda_p\right]\,{v^2_{\Phi}}\;.
\end{align}%
\end{subequations}%
This shows their nature as dilaton as well as pNGB because it corresponds to \mbox{$m^2_{h_\Phi}\approx\beta_{\lambda_\Phi} v^2_\Phi$} and \mbox{$m^2_h\approx2\left(\lambda_\Phi\beta_{\lambda_p}/\beta_{\lambda_\Phi}-\lambda_p\right)v^2_\Phi$}, where $\beta_i$ are the respective beta function coefficients, given in App.~\ref{app:beta_functions}. $\lambda_H$~remains close to its SM value and the EW scale VEV gets to keep the SM relation
\begin{equation}\label{eq:SMvevMass}
v_H^2\approx \frac{m^2_h}{2\lambda_H}\;.
\end{equation}
The EW scale is, therefore, custodially suppressed compared to the intermediate scale~$v_\Phi$ of spontaneous scale and custodial symmetry violation. 

\section{Minimal model}\enlargethispage{3cm}
\renewcommand{\arraystretch}{1.15}
\begin{table}[t]
	\begin{tabular}{ccccc}
		Name & \#Gens.  & $\SU{3}_\mathrm{c}\!\times\!\SU{2}_\mathrm{L}\!\times\!\U{1}_\mathrm{Y}$\!&\!$\times\U{1}_\mathrm{X}$\! & ~~~~$\U{1}_\mathrm{B-L}$ \\ 
		\hline 
		$Q$ & $3$  & $\left(\rep{3},\rep{2},+\frac16\right)$  & $-\frac{2}{3}$ & ~~~~$+\frac{1}{3}$ \\
		$u_R$ & $3$  & $\left(\rep{3},\rep{1},+\frac{2}{3}\right)$  & $+\frac{1}{3}$ & ~~~~$+\frac{1}{3}$ \\
		$d_R$ & $3$ & $\left(\rep{3},\rep{1},-\frac{1}{3}\right)$ & $-\frac{5}{3}$ & ~~~~$+\frac{1}{3}$ \\
		$L$ & $3$ & $\left(\rep{1},\rep{2},-\frac12\right)$ & $+2$ & ~~~~$-1$ \\
		$e_R$ & $3$ & $\left(\rep{1},\rep{1},-1\right)$  & $+1$ & ~~~~$-1$ \\
		$\nu_R$ & $3$ & $\left(\rep{1},\rep{1},\phantom{-}0\right)$ & $+3$ & ~~~~$-1$ \\
        \hline
		$H$ & $1$ & $\left(\rep{1},\rep{2},+\frac12\right)$ & $+1$ & ~~~~$\phantom{-}0$ \\
		$\Phi$ & $1$ & $\left(\rep{1},\rep{1},\phantom{-}0\right)$ & $+1$ & ~~~~$-\frac{1}{3}$ \\[1pt]
		\hline
	\end{tabular}
	\caption{\label{tab:model}
	Field content of a minimal model that can realize the idea of ``Custodial Naturalness.'' $\U{1}_\mathrm{X}$ is a new gauge symmetry and the last column shows the linear combination of $\U{1}_\mathrm{X}$ and $\U{1}_\mathrm{Y}$ to $\U{1}_\mathrm{B-L}$. The $\U{1}_\mathrm{X}$ charges of fermions are entirely determined by the choice $q_\Phi^{\mathrm{B-L}}=-\frac{1}{3}$. Other choices are possible (see footnote~\ref{fn:charge}) and may be required in a UV embedding.}
\end{table}
The minimal extension of the SM (incl.~$3$ generations of right-handed neutrinos) to realize our idea of ``Custodial Naturalness'' is displayed in Tab.~\ref{tab:model}. $H$ and the new complex scalar $\Phi$ are charged under a new, family universal $\U{1}$ gauge group. Anomaly cancellation requires the charges of the SM fermions under the new $\U{1}$ to be linear combinations of their hypercharge and $B-L$. Hence, $\Phi$ has a $B-L$ charge which we denote as $q^{\mathrm{B-L}}_\Phi$.\footnote{%
Our model is very similar to the classical conformal extension~\cite{Iso:2009ss,Iso:2009nw} of the ``minimal $B-L$ model''~\cite{Khalil:2006yi} (see also~\cite{Davidson:1978pm,*Marshak:1979fm,*Mohapatra:1980qe,*Wetterich:1981bx,*Jenkins:1987ue,*Buchmuller:1991ce}). As a main difference, we include the assumption of custodial symmetry and take into account gauge kinetic mixing which excludes the charge assignment $q^{\mathrm{B-L}}_\Phi=2$ of~\cite{Iso:2009ss,Iso:2009nw} for our purpose, see discussion in footnote~\ref{fn:charge}.}

It is always possible to choose a linear combination of $\U{1}$'s such that the  charge assignment of $H$ and $\Phi$ is symmetric. We call this linear combination $\U{1}_\mathrm{X}$,
\begin{equation}
\vspace{-0.1cm}
    Q^{(\mathrm{X})}~=~2\,Q^{(\mathrm{Y})}+\frac{1}{q^{\mathrm{B-L}}_\Phi}\,Q^{(\mathrm{B-L})}\;.
\vspace{-0.1cm}
\end{equation}
The charge of $\Phi$ is the only free parameter of the charge assignment and we fix it to $q^{\mathrm{B-L}}_\Phi=-1/3$ for the sake of this letter. The corresponding charges of scalars and fermions are shown in Tab.~\ref{tab:model}.\footnote{\label{fn:charge}%
Possible charge assignments are roughly bounded by $1/3\lesssim |q^{\mathrm{B-L}}_\Phi|\lesssim 5/11$ for custodial symmetry violation through gauge kinetic mixing to remain small. If $\left.g_{12}\right|_{M_\mathrm{Pl}}=0$ is imposed, one needs $|q^{\mathrm{B-L}}_\Phi|<3/8$ to trigger EWSB. For the special value \mbox{$q^{\mathrm{B-L}}_\Phi=-\frac{16}{41}$} gauge kinetic mixing is absent at one loop. (This
value was independently found by~\cite{Oda:2015gna,Das:2016zue}, while the corresponding condition is known as ``charge orthogonality''~\cite{Loinaz:1999qh}, see also~\cite{Carone:1995pu,Chang:2000xy, Aranda:2000ma}).}

\subsection{The effective potential}
The effective potential for background fields $H_b$ and $\Phi_b$ at one loop in $\overline{\text{MS}}$ is given by
\begin{equation}\label{eq:V_eff}%
\hspace{-0.2cm}
V_\text{eff}=V_\text{tree}+ \sum_i \frac{n_i (-1)^{2 s_i}}{64\pi^2}m_{i,\text{eff}}^4\left[\ln\left(\frac{m_{i,\text{eff}}^2}{\mu^2}\right)-C_i\right]%
\end{equation}
where $(-1)^{2s_i}$ is $\varpm1$ for bosons(fermions), $n_i$ is the number of degrees of freedom and $C_i=\frac{5}{6}\!\!\left(\frac{3}{2}\right)$ for vector bosons(scalars and fermions).

The vector boson effective masses are given by \mbox{$m_W^2=\frac{1}{2}g_L^2H_b^2$} and (see App.~\ref{app:mass_diagonalization})
\begin{subequations}\label{eq:mV}
\begin{align}
        &m_Z^2=\frac{1}{2}(g_L^2+g_Y^2)H_b^2-\mathcal{O}\left(H_b^4/\Phi_b^2\right),& \\
        \raisetag{17pt}\label{eq:mZp}
        &m_{Z'}^2=2 g_X^2\Phi_b^2+\frac{1}{2}(g_{12}+2g_X)^2H_b^2+\mathcal{O}\left(H_b^4/\Phi_b^2\right).&
\end{align}%
\end{subequations}%
The effective scalar masses are the eigenvalues of $(m_\text{eff}^2)_{a,b}\equiv\partial_{\phi_a}\partial_{\phi_b} V_\text{tree}$, and for the top quark \mbox{$m_t=y_tH_b$}.

To show analytically how custodial symmetry violation affects the Higgs potential, we define a new potential
\begin{align}
\vspace{-0.1cm}
V_\mathrm{EFT}(H_b)~:=~V_\mathrm{eff}\left(H_b,\tilde{\Phi}(H_b)\right)\;,
\vspace{-0.1cm}
\end{align}
where $\tilde{\Phi}(H_b)$ is implicitly defined by the constraint\footnote{%
This resembles effective field theory methods~\cite{Burgess:2007pt,Manohar:2020nzp}.} 
\begin{align} 
\vspace{-0.1cm}
	\left.\frac{\partial V_\text{eff}}{\partial \Phi_b}\right|_{\Phi_b=\tilde{\Phi}(H_b)}=0\;.
\vspace{-0.1cm}
\end{align}
$V_\text{EFT}$ is purely a function of $H_b$ and has a minimum at $H_b=\langle H\rangle=v_H/\sqrt{2}$ just as the original effective potential. We expand $V_\text{EFT}$ around its minimum using \mbox{$H_b\ll\tilde\Phi (H_b/\Phi_b=0)=:\Phi_0$}. $\Phi_0$ is to a good approximation the VEV of $\Phi$, given by the usual result
\begin{align}\label{eq:Phi0_short}
\vspace{-0.1cm}
    \Phi_0^2\approx\exp\left\{-\frac{16\pi^2\lambda_\Phi}{3g_X^4}-\ln(2g_X^2)+\frac13+...\right\}\mu^2\;.
\vspace{-0.1cm}
\end{align}
At the quadratic order in $H_b$,
\begin{align}\notag
    V_\text{EFT}\approx2\left[\lambda_p-\left(1+\frac{g_{12}}{2\,g_X}\right)^2\lambda_\Phi\right]\Phi_0^2H_b^2+\frac{\lambda_p\lambda_H}{16\pi^2}[...]\;.
\end{align}
This expression is RG-scale independent and shows how custodial symmetry violating terms generate the Higgs quadratic term via the differential running of \mbox{$\lambda_\Phi-\lambda_p$}.

A different expansion is useful to understand the matching to
the SM. We take $\mu\sim\langle\Phi\rangle$ to avoid large logarithms. A particularly convenient choice\footnote{%
None of our conclusions depend on this choice. 
}
is $\mu=\mu_0:=\sqrt{2} g_X\Phi_0 \mathrm{e}^{-1/6}$. 
At $\mu=\mu_0$ we perform a 't Hooft-like expansion of $V_\mathrm{eff}$, assuming
\begin{equation}\label{eq:epsilon}
\vspace{-0.1cm}
\frac{\lambda_p}{\lambda_H}\sim\frac{H_b^2}{\Phi_0^2}\sim\epsilon^2\rightarrow0\;,
\vspace{-0.1cm}
\end{equation}
sending $\epsilon\rightarrow0$ while keeping $\lambda_p\Phi^2_0\sim\lambda_HH^2_b$ fixed.\footnote{\label{fn:expansion}%
The quantitative difference between the ``$H_b\ll\Phi_0$'' and ``$\epsilon$'' expansions is of $\mathcal{O}(\lambda_p^2)$ $\left[\mathcal{O}(\lambda_p)\right]$ for the Higgs mass [quartic] term.}
This is consistent with Gildener-Weinberg conditions~\cite{Gildener:1976ih},
which yield $\lambda_p\Phi^2_b\sim\lambda_HH^2_b$ as the flat direction 
of the tree level potential at the RG-scale where $\lambda_\Phi=\lambda_p^2/\lambda_H$~\cite{Sher:1988mj}.
The potential up to $\epsilon^4$ can be written as
\begin{widetext}
\begin{equation}\label{eq:V_EFT conf}
\vspace{-0.1cm}
	V_\text{EFT}=-\frac{6\,g_X^4}{64\pi^2}\Phi_0^4+2\,\lambda_p\Phi_0^2H_b^2+\lambda_HH_b^4+\sum_i\frac{n_i (-1)^{2s_i}}{64\pi^2}m_{i,\text{eff}}^4\left[\ln\left(\frac{m_{i,\text{eff}}^2}{\mu_0^2}\right)-C_i\right],
\vspace{-0.1cm}
\end{equation}
\end{widetext}\enlargethispage{1.8cm}
where the sum now only runs over the effective masses of the SM with tree level potential \mbox{$V_\text{tree}=2\lambda_p\Phi_0^2|H|^2+\lambda_H|H|^4$}. 
Importantly, Eq.~\eqref{eq:V_EFT conf} agrees with the SM effective Higgs potential at this scale~\cite{Martin:2001vx}. Therefore, we chose $\mu=\mu_0$ as matching scale to the SM, implying that $\left.\lambda_p\Phi_0^2\right|_{\mu_0}$ and $\left.\lambda_H\right|_{\mu_0}$ are determined by the SM Higgs mass term and quartic coupling.

We stress that the above expansions are displayed only for intuition and sanity checks. Our quantitative analysis is based on a fully numerical evaluation of the effective potential~\eqref{eq:V_eff}.
\vspace{-0.7cm}

\subsection{Masses and mixing}
While tree level expressions, Eq.~\eqref{eq:mV} and App.~\ref{app:mass_diagonalization}, are good approximations for the vector boson masses, 
reliable expressions for scalar masses and mixings require minimization of the full effective potential.
The exact expressions are not instructive but reasonably well approximated by Eq.~\eqref{eq:mhPhi}. 
The Higgs-dilaton mixing angle is approximately 
\begin{equation}
    \tan\theta\approx\frac{2\left[\lambda_p-\left(1+\frac{g_{12}}{2g_X}\right)^2\left(\lambda_\Phi-\frac{3g_X^4}{16\pi^2}\right)\right]v_H v_\Phi}{m_h^2-m_{h_\Phi}^2}\;.
\end{equation}
This induces a small dilaton coupling to the SM via operators $\mathcal{O}_{h_\Phi}\approx \sin\theta \times \mathcal{O}^{\mathrm{SM}}_{h\rightarrow h_\Phi}$ that contain insertions of $h_\Phi$ instead of $h$. In addition, the dilaton couples to SM fields via the trace anomaly with generically suppressed couplings $\propto h_{\Phi}/v_{\Phi}$, see e.g.~\cite{Goldberger:2007zk,Chacko:2012sy,Bellazzini:2012vz}.

We use a fully numerical evaluation of all masses and mixings for our analysis which also confirms the analytic approximations.

\section{Numerical analysis}
Fixing the observables $G_\mathrm{F}$, $m_h$, and $m_t$ in terms of the free parameters $\lambda$, $g_X$ and $y_t$ (fixed at $M_\mathrm{Pl}$ with $g_{12}=0$)\footnote{This condition is justified by increasing custodial symmetry.} there are no additional degrees of freedom. Hence,  $m_{h_\Phi}$ and $m_{Z'}$ (incl. their couplings, production cross sections and branching ratios) are predictions of the model. Correlations of predictions can be relaxed by allowing for $g_{12}\neq0$ at $M_\mathrm{Pl}$, or more generally, by additional new sources of custodial symmetry breaking~\cite{DeBoer:2024xx}.

\begin{table*}
\renewcommand{\arraystretch}{1.25}
	\centering
	\begin{tabular}{ccccccccccc}\hline
    ~$\mu\left[\mathrm{GeV}\right]$ ~&~ $g_X$ ~& $g_{12}$ &~ $\lambda_H$ ~&~ $\lambda_p$ ~&~ $\lambda_\Phi$ ~& $y_t$ &~ $m_{h_\Phi}\left[\mathrm{GeV}\right]$ ~&~ $m_{Z'}\left[\mathrm{GeV}\right]$    ~&~
    $m_{h}\left[\mathrm{GeV}\right]$ ~&~
    $v_H\left[\mathrm{GeV}\right]$
    \\\hline
    $1.2\cdot10^{19}$ ~&~ $0.0713$ ~& $0.$ & \multicolumn{3}{c}{$\lambda_H=\lambda_p=\lambda_\Phi=3.0304\cdot10^{-5}$} & $0.377$ &~ - ~&~ - ~&~ - ~&~ - \\
    $4353$ ~&~ $0.0668$ ~& $0.0093$ &~ $\bf{0.084}$ ~&~ $-1.6\cdot10^{-6}$ ~&~ $2.5\cdot10^{-11}$ ~& $\bf{0.795}$ &~ $67.0$ ~&~ $5143$ ~&~ $\bf{132.0}$ ~&~ $\bf{263.0}$\\ 
    $172$ ~&~ - ~&~ - ~&~ $0.13$ ~&~ - ~&~ - ~& $0.930$ &~ - ~&~ - ~&~ $125.3$ ~&~ $246.1$\\
    \hline
	\end{tabular}
 \caption{\label{tab:benchmark}%
 Input parameters of an example benchmark point (BP) at the high scale (top) and corresponding predictions at the matching scale $\mu_0$~(middle) and $m_t$ (bottom). At $\mu_0$ the bold parameters also correspond to the parameters of the one-loop SM effective potential, Eq.~\eqref{eq:V_EFT conf}. The numerical result for the VEV of $\Phi$ is $\langle\Phi\rangle=v_\Phi/\sqrt{2}=54407\,\mathrm{GeV}$.
 } 
\end{table*}
To explore the parameter space we perform a scan. Our algorithm for finding a reasonable range of starting parameters is described in App.~\ref{app:numeric}. At the high scale, we impose $\SO{6}$ symmetric boundary conditions $\left.\lambda_{H,\Phi,p}\right|_{M_\mathrm{Pl}}=\left.\lambda\right|_{M_\mathrm{Pl}}$ and $\left.g_{12}\right|_{M_\mathrm{Pl}}=0$. We iteratively use \eqref{eq:Phi0_long} to determine $\Phi_0$ and $\mu_0$, then 2-loop evolve the model down to $\mu_0$. 
At $\mu_0$ we numerically minimize the 1-loop effective potential to compute the VEVs $v_\Phi$, $v_H$, scalar masses and couplings. These are predictions of each model point and we can match them to the according parameters of the SM 1-loop effective potential.
From $\mu_0$ we 2-loop evolve the SM to $m_t$ (dilaton contributions are negligible). We sort out points that do not reproduce the experimentally determined EW scale within \mbox{$v^{\mathrm{exp}}_H=246.2\pm0.1\,\mathrm{GeV}$}\footnote{%
\label{foot:G_F}%
We remark that the EW scale is known to better precision from measurement of $G_\mathrm{F}$. Nonetheless, we allow for a larger error here as to allow for more parameter points to pass this selection. There are no significant changes in our analysis under a variation of this window.} 
or disagree with the SM values of $g_L$, $g_Y$, $g_3$ or $y_t$ including errors. 
The Higgs mass has to stay in its experimentally determined $3\sigma$ range $m^{\mathrm{exp}}_h=125.25\pm0.51\,\mathrm{GeV}$~\cite{ParticleDataGroup:2022pth} for a point to be viable but we do not always enforce this constraint
in order to investigate correlations. The new couplings $g_{X}$, $g_{12}$ and masses $m_{Z'}$, $m_{h_\Phi}$ at the low scale are predictions of each model point.

\begin{figure*}[!t!]
\includegraphics[width=0.48\linewidth]{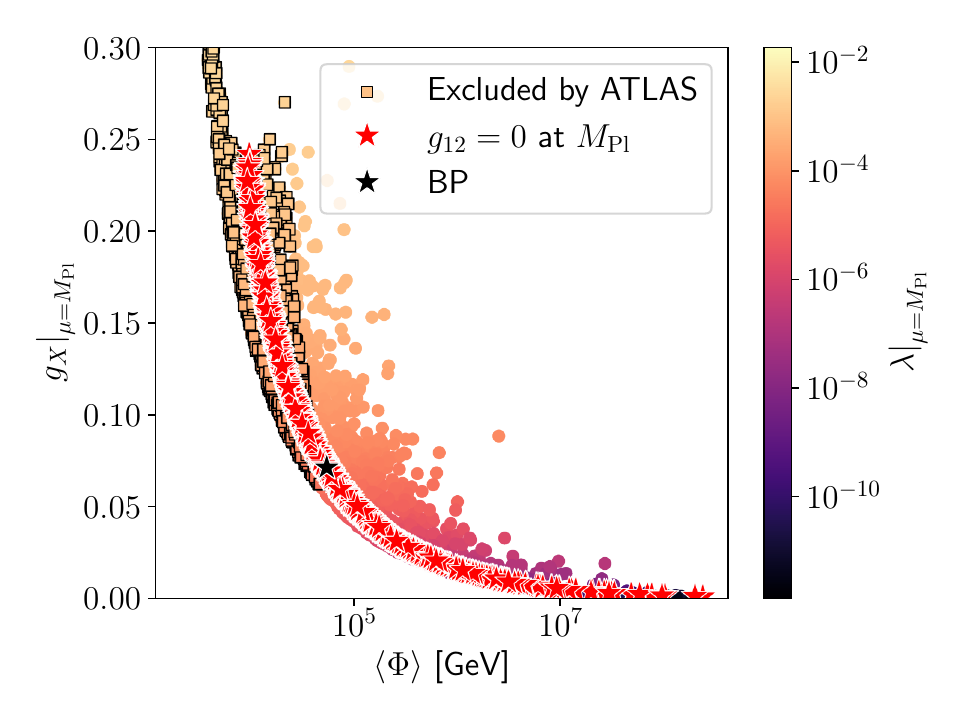}%
\hfill
\includegraphics[width=0.48\linewidth]{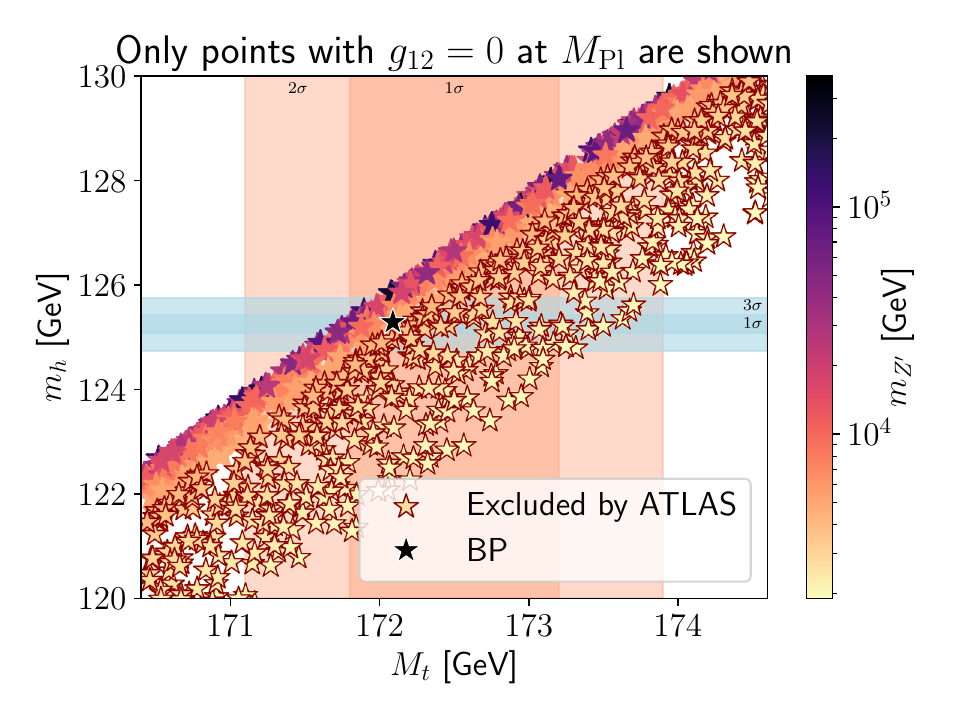}
\caption{\label{fig:scan}%
Left: Parameters of our model (at the scale $\mu=M_\mathrm{Pl}$) and 
corresponding prediction of the new scale $\langle\Phi\rangle$. 
Points with red stars obey $\left.g_{12}\right|_{M_\mathrm{Pl}}=0$. 
Right: Correlation of the predictions of $M_t$ (top pole mass), $m_h$ and $m_{Z'}$ for all points with $\left.g_{12}\right|_{M_\mathrm{Pl}}=0$. All points shown reproduce the correct EW scale.}
\end{figure*}

\begin{figure}
 	\centering
	\includegraphics[width=0.48\textwidth]{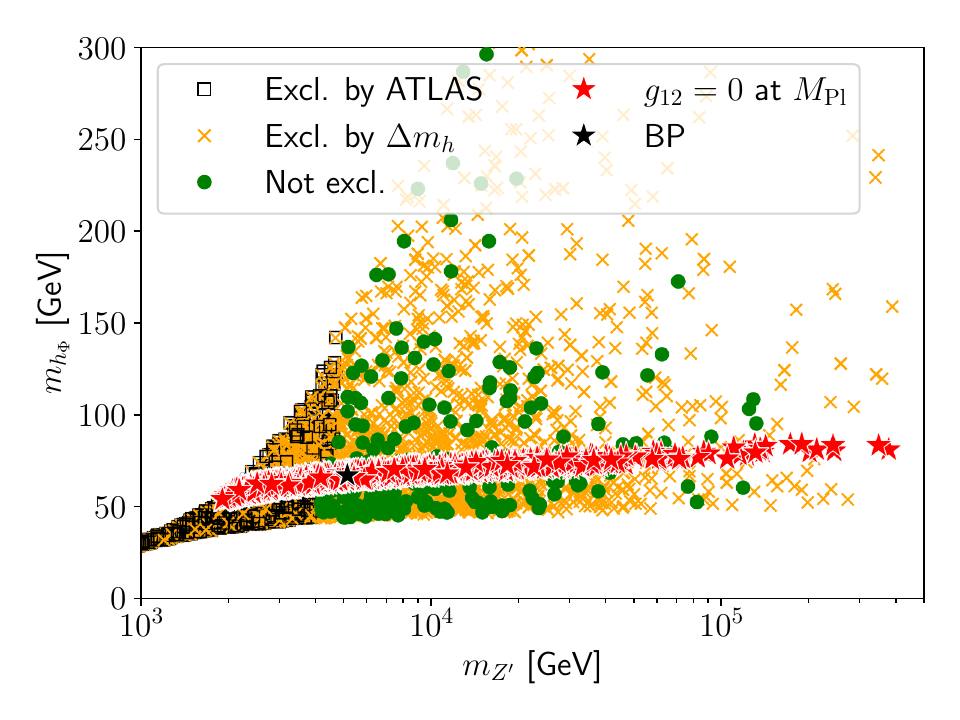}%
 	\caption{\label{fig:correlation}%
  Correlation of the predictions of the new particle masses $m_{Z'}$ and $m_{h_\Phi}$.}
 \end{figure}

In Fig.~\ref{fig:scan} (left) we show points that reproduce the correct EW scale, which leads to a strong correlation of the permitted $g_X$ and $\lambda$. Points marked by red stars are most predictive because we require $\left.g_{12}\right|_{M_\mathrm{Pl}}=0$. Other points have a random value $\left.g_{12}/{g_X}\right|_{M_\mathrm{Pl}}\in[-0.1,0.1]$ to demonstrate the widening of the parameter space in the presence of additional sources of custodial breaking. The correlation of $m_h$, $M_t$ (top pole mass) and $m_{Z'}$ -- exclusively in the case with $\left.g_{12}\right|_{M_\mathrm{Pl}}=0$ -- is shown in Fig.~\ref{fig:scan} (right). Fig.~\ref{fig:correlation} shows the corresponding predictions for $m_{Z'}$ and $m_{h_\Phi}$ and their correlation. For $\left.g_{12}\right|_{M_\mathrm{Pl}}=0$ the dilaton mass is always smaller than the Higgs mass.  
We show a benchmark point (BP) of our model in Tab.~\ref{tab:benchmark} and as a black star in the figures. \enlargethispage{2cm}

\section{Phenomenological constraints}
\begin{figure}[t]
	\centering
	\includegraphics[width=0.5\textwidth]{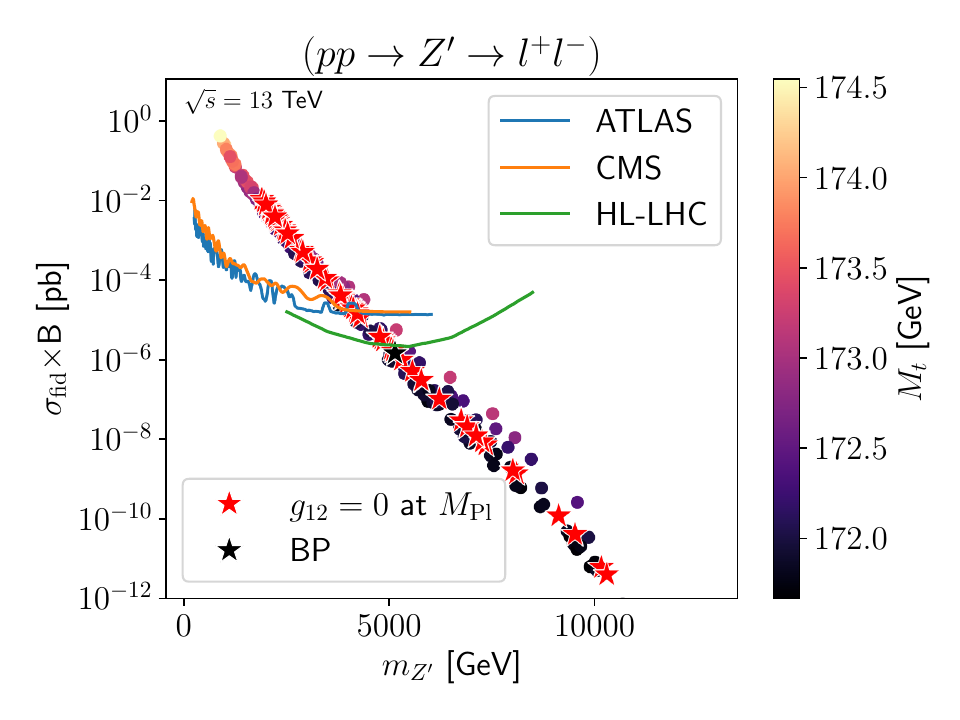}
	\caption{\label{fig:dilepton}
	Parameter points and $95\%\,\mathrm{C.L.}$ exclusion contours for $m_{Z'}$ from ATLAS and CMS dilepton resonance searches~\cite{ATLAS:2019erb,CMS:2021ctt} (using fiducial cuts from~\cite{ATLAS:2019erb}) and projections for HL-LHC (at $14\,\mathrm{TeV}$). All points reproduce the correct EW scale, Higgs and top mass.}
\end{figure}

Given the charges of Tab.~\ref{tab:model}, the $Z'$ production cross section and branching ratios are predictions of the model. We compute them using \texttt{MadGraph5\_aMC@NLO}~\cite{Alwall:2014hca} with a \texttt{UFO} input~\cite{Degrande:2011ua} generated with \texttt{FeynRules}~\cite{Christensen:2008py}.  
Dilepton resonance searches~\cite{ATLAS:2019erb,CMS:2021ctt} are the most important constraint on our model and already exclude $m_{Z'}\lesssim4\,\mathrm{TeV}$, see Fig.~\ref{fig:dilepton}. Points excluded by ATLAS~\cite{ATLAS:2019erb} are marked on all of our plots, while CMS gives similar limits~\cite{CMS:2021ctt}.

New sources of custodial symmetry violation lead to a shift of the $Z$ mass $\Delta m_Z\approx-m_Z \langle H\rangle^2/(2\langle\Phi\rangle^2)$, see Eq.~\eqref{eq:mZp} and App.~\ref{eq:mZ}. With all couplings at their SM values, $m_Z$ stays within its $2\sigma$ uncertainty~\cite{ParticleDataGroup:2022pth} if $\langle\Phi\rangle\gtrsim18\,\mathrm{TeV}$. 
Direct searches supersede this constraint which explains our simplified treatment.
 
The dilaton is heavier than $m_h-m_Z$ but could be lighter than half the Higgs mass~(see Fig.~\ref{fig:correlation}). The new invisible Higgs decay $h\rightarrow h_\Phi h_\Phi$ is highly suppressed.
Constraints on a general dilaton are given in~\cite{Ahmed:2015uqt,Ahmed:2019csf} but are easily avoided due to the large value of $v_\Phi$.
Assuming dilaton couplings to the SM scale with the Higgs mixing, a naive rescaling of rates suggests that for $\sin^2\theta\sim\mathcal{O}(10^{-5})$ (as our BP) we could produce about one dilaton per $10^{5}$ Higgses. 
The corresponding dilaton lifetime of $\tau_{h_\Phi\rightarrow\mathrm{SM}}\sim\mathcal{O}(10^{-15}\,\mathrm{s})$
would require $\mu\mathrm{m}$ vertex tracker resolution or corresponding initial state boosts to yield a detectable displaced vertex signature at a Higgs factory.

\section{Fine tuning}
\begin{figure}
 	\centering
	\includegraphics[width=0.5\textwidth]{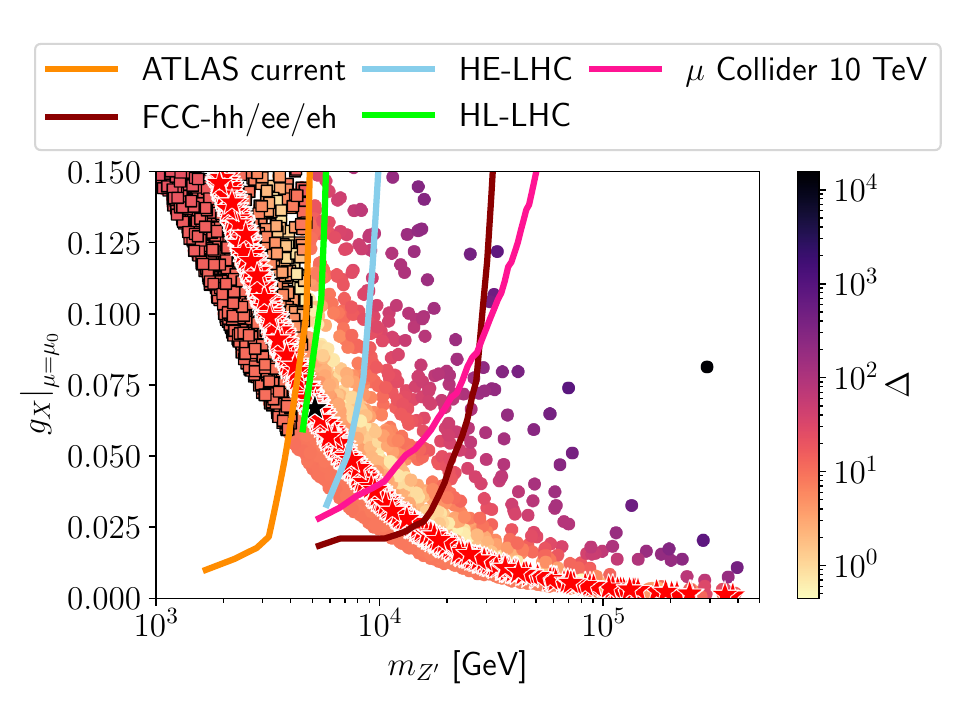}%
 	\caption{\label{fig:fine_tuning}%
  Heavy vector mass $m_{Z'}$ vs.\ coupling at the SM matching scale $\left.g_X\right|_{\mu_0}$ and fine tuning of all valid parameter points (red stars have $\left.g_{12}\right|_{M_\mathrm{Pl}}=0$). 
We overlay current constrains from ATLAS~\cite{ATLAS:2019erb} and projections (for a hypercharge universal $Z'$) for future colliders extracted from~\cite{EuropeanStrategyforParticlePhysicsPreparatoryGroup:2019qin}.}
\end{figure}

To demonstrate the absence of the little hierarchy problem we quantify the amount of fine tuning necessary to generate a hierarchy between $\langle H\rangle$ and $\langle\Phi\rangle$. Inspired by Barbieri-Giudice~\cite{Barbieri:1987fn}, for any coupling $g_i$ we define 
\begin{align}
 \Delta_{g_i}^{\left(\frac{\langle H\rangle}{\langle\Phi\rangle}\right)}:=\,\left|\frac{
 \partial\,\ln\frac{\langle H\rangle}{\langle \Phi\rangle}}{\partial\ln g_i}\right|\,=\,\left|\frac{g_i}{\langle H\rangle}\frac{\partial\langle H\rangle}{\partial g_i}-\frac{g_i}{\langle\Phi\rangle}\frac{\partial\langle \Phi\rangle}{\partial g_i}\right|.
\end{align}
This choice subtracts the shared sensitivity of VEVs to variations of $g_i$ which is meaningless in scenarios of dimensional transmutation~\cite{Anderson:1994dz}. \enlargethispage{2cm} Fine tuning of a given point in parameter space is then quantified as 
\begin{equation}\label{eq:fine_tuning}
	\Delta~:=~\max_{g_i}\;\Delta_{g_i}^{\left(\frac{\langle H\rangle}{\langle\Phi\rangle}\right)}~=~\max_{g_i}\;\left|\Delta_{g_i}^{\langle H\rangle}-\Delta_{g_i}^{\langle\Phi\rangle}\right|.
\end{equation}
The fine tuning of parameter points is shown in Fig.~\ref{fig:fine_tuning}. Most points have $\Delta\lesssim10$, i.e.\ do not require tuning while allowing for a hierarchy of $\langle H\rangle\approx10^{-3}\langle\Phi\rangle$ consistent with custodial suppression.

\section{Extensions and Embeddings}\enlargethispage{1cm}
The mechanism of ``Custodial Naturalness'' is reasonably stable under variations of 
boundary conditions, charge assignments, or the addition of extra particles~\cite{DeBoer:2024xx}.
Minimal fermionic extensions can populate the neutrino portal and provide ingredients for neutrino mass generation~\cite{Foot:2007ay,Iso:2009ss} or an explanation of Dark Matter (DM)~\cite{Okada:2018ktp}. Without extra symmetries, our dilaton has a short lifetime and cannot be the DM, cf.~\cite{Chang:2007ki, Alanne:2014kea}.

New fermions would help to avoid the vacuum instability when $\lambda_H$ runs negative, 
which in the minimal model happens around $10^{15}-10^{17}\,\mathrm{GeV}$~(see also~\cite{Oda:2015gna,Das:2015nwk,Das:2016zue}). $\lambda_H$ crosses back to positive values at an even higher scale, implying that $\Lambda_\text{high}$ 
can be either below or above the $\lambda_H<0$ region. The benchmark point of Tab.~\ref{tab:benchmark} realizes the latter possibility, see Fig~\ref{fig:running}. 

We stress that a small value of $\left.\lambda_H\right|_{\Lambda_\text{high}}$ is required, as larger values of $\lambda_{p,\Phi}$ would require a larger value of $g_X$ which in turn brings the $\U{1}_\mathrm{X}$ Landau pole down to $M_\mathrm{Pl}$. 
Our model showcases the possible connection of the EW hierarchy problem to the phenomenon of quantum criticality: Custodial symmetry could originate from an interacting UV fixed point, see e.g.~\cite{Litim:2014uca}, at a quantum critical value of the scalar couplings close to the scale of a putative zero crossing of $\lambda_H$.

Extensions are required for an embedding into a grand unified theory (GUT).
The size of gauge kinetic mixing at the high scale may be calculable if the custodial group is embedded as a subgroup
$G_\mathrm{cust.}\subset G_{\mathrm{GUT}}$ just like for the SM $\SO{4}\subset\SO{10}$ in Pati-Salam unification~\cite{Pati:1974yy}. The custodial naturalness mechanism of scale separation could then also be used to explain doublet-triplet splitting in the fashion of Dimopoulos-Wilczek~\cite{Dimopoulos:1981xm,Babu:1993we}.

In the future one should also include finite temperature effects which we have ignored here. The fact that the CW phase transition is generically first order~\cite{Linde:1975sw,Weinberg:1976pe,Litim:1994jd} 
naturally invites investigations of baryo- or leptogenesis in our model and extensions~\cite{Iso:2010mv,Das:2024gua,Chauhan:2024jfq}, and may also give rise to observable gravitational wave signals~\cite{Jinno:2016knw,Marzola:2017jzl,Prokopec:2018tnq,Marzo:2018nov,Ellis:2020nnr,Dasgupta:2022isg,Huang:2022vkf}. 

\section{Conclusions}\enlargethispage{1cm}
We have presented a new idea to address the large separation between $M_\mathrm{Pl}$ and the EW scale by classical scale invariance, paired with an enhanced custodial symmetry. Earlier works have already used dimensional transmutation in a hidden sector to generate the EW scale via portals, which often necessitates tuning in the form of a little hierarchy problem. 
Here, we emphasize the importance of custodial symmetry and its spontaneous breaking to get around this. We also stress the importance of taking into account all possible portals, including gauge kinetic mixing, which also modifies the dynamics of the Higgs portal. The spontaneous breaking of scale symmetry in our case also leads to the spontaneous breaking of custodial symmetry. The Higgs boson is identified with a pNGB of custodial symmetry which naturally explains its suppressed mass and the suppression of the EW scale without a little hierarchy problem. 

In the minimal realization discussed here, the SM is extended by a single complex scalar field and a new $\U{1}_\mathrm{X}$ gauge symmetry. In addition, we assume classical scale invariance and $\SO{6}$ custodial symmetry at a high scale. Due to the enhanced symmetry, the number of free parameters stays the same as in the SM which makes our model predictive. The most prominent experimental signature is a new $Z'$ gauge boson which is a well motivated target for resonance searches at future colliders, see Fig.~\ref{fig:fine_tuning}. The model also predicts a dilaton with mass $m_{h_\Phi}\sim75\,\mathrm{GeV}$, fixed couplings to the SM, and a potentially long enough lifetime for it to be a target for displaced vertex searches at future Higgs factories. Another prediction is that the top pole mass should be constrained to the lower end of its presently allowed $1\sigma$ mass range. 

The basic mechanism underlying our idea is stable under extensions of the simplest model. Such extensions can include mechanisms for the generation of neutrino masses, DM, and the baryon asymmetry of the Universe and this shall be investigated in the future.

\vspace{-0.5cm}
\acknowledgments
\vspace{-0.2cm}
We are grateful to Florian Goertz for helpful discussions on composite Higgs models and fine tuning, as well as to Aqeel Ahmed and Jisuke Kubo for very useful discussions and insightful comments on the manuscript.

\onecolumngrid
\appendix
\section{Beta functions}\label{app:beta_functions}
We neglect all Yukawa couplings besides $y_t$ and allow for general $\U{1}_\mathrm{X}$ charges $q_{H,\Phi}$ of the Higgs and $\Phi$. The beta functions at one loop are given by
\begin{equation}
\begin{split}
\beta_{\lambda_H} ~=~&\frac{1}{16\pi^2}\biggl[
+ \frac{3}{2} \left(\left(\frac{g_Y^{2}}{2}+\frac{g_L^2}{2}\right)+2\left(q_Hg_{X}+\frac{g_{12}}{2}\right)^{2}\right)^2+\frac{6}{8}g_L^4 - 6 y_t^4\\&\hspace{6ex}
+ 24 \lambda_H^{2} + 4\lambda_p^{2} + \lambda_H\left( 12  y_t^2 - 3 g_Y^{2} - 12\left(q_H g_{X}+\frac{g_{12}}{2}\right)^{2} - 9 g_L^{2} \right)\biggl]\;,\\
\beta_{\lambda_\Phi} ~=~&\frac{1}{16\pi^2}\left(
+ 6 q_\Phi^4g_{X}^{4}
+ 20 \lambda_\Phi^{2}
+ 8 \lambda_p^{2}
- 12 \lambda_\Phi q_\Phi^2 g_{X}^{2} 
\right)\;,\\
\beta_{\lambda_p} ~=~&\frac{1}{16\pi^2}\biggl[
+ 6 q_\Phi^2 g_X^2\left(q_Hg_X+\frac{g_{12}}{2}\right)^2 + 8 \lambda_p^{2}\\&\hspace{6ex}
+\lambda_p \left(8 \lambda_\Phi + 12 \lambda_H -  \frac{3}{2} g_Y^{2} - 6q_\Phi^2g_X^2-6\left(q_Hg_X+\frac{g_{12}}{2}\right)^2 - \frac{9}{2} g_L^{2}+ 6 y_t^2\right)\biggr]\;, \\
\beta_{g_{12}}~=~& \frac{1}{16\pi^2}\left[-\frac{14}{3}g_X g_Y^2-\frac{14}{3}g_Xg_{12}^2+\frac{41}{3}g_Y^2 g_{12}+\frac{179}{3}g_X^2g_{12}+\frac{41}{6}g_{12}^3\right]\;.
\end{split}
\end{equation}
For the charges shown in Tab.~\ref{tab:model} the dominant splitting of $\lambda_\Phi-\lambda_p$ via running is given by 
\begin{equation}
    \beta_{\lambda_\Phi}-\beta_{\lambda_p}=-\frac{6\,g_{12}\,g_X^2}{16\pi^2}\left(g_X+\frac{g_{12}}{4}\right)-\frac{\lambda_p}{16\pi^2}\left[6 y_t^2-\frac92 g_L^2-\frac32 g_Y^2+12 (\lambda_H-\lambda_p) \right]+\dots\;,
\end{equation}
where we have dropped higher powers of small parameters. For our parameters, the dominant source of $\lambda_\Phi-\lambda_p$ splitting is $g_{12}$. The second term is subdominant here because criticality of the CW mechanism ($\beta_{\lambda_\Phi}\approx\lambda_\Phi$) with $g_X\approx0.1$ requires a small $\lambda_\Phi$, hence small $\lambda_H=\lambda_p=\lambda_\Phi\lesssim10^{-4}$ at the high scale. We remark that we perform the numerical running with the full two-loop beta functions computed with \texttt{PyR@TE}~\cite{Sartore:2020gou}.

\section{Tree level effective masses}
\label{app:mass_diagonalization}
For background fields $H=H_b$, $\Phi=\Phi_b$ the tree level effective mass matrix of the neutral vector bosons is given by
\begin{equation}\label{eq. vector effective masses}
	M_V=\left(\begin{matrix}
		\frac{g_Y^2}{2} H_b^2 &-\frac{g_Y g_L}{2} H_b^2 & \frac{\left( 2g_X+g_{12}\right)g_Y}{2} H_b^2\\
    	-\frac{g_Y g_L}{2}H_b^2 & \frac{g_L^2}{2} H_b^2 &-\frac{\left( 2g_X+g_{12}\right)g_L}{2}H_b^2 \\
    	\frac{\left( 2g_X+g_{12}\right)g_Y}{2} H_b^2 &-\frac{\left( 2g_X+g_{12}\right)g_L}{2} H_b^2 & 2\left(\frac{2g_X+g_{12}}{2}\right)^2\!\!H_b^2+2g_X^2\Phi_b^2	\end{matrix}\right).
\end{equation}	
$M_V$ can be diagonalized by two consecutive orthogonal rotations, combined into a mixing matrix as
\begin{equation}
    U=\left(\begin{matrix}c&-s c'&s s'\\s&c c'&-c s'\\0&s'&c'\end{matrix}\right)
\end{equation}
where $s=\sin\theta_W$, $c=\cos\theta_W$ with the EW mixing angle $\tan\theta_W:= g_Y/g_L$, while $s'=\sin\theta'$ and $c'=\cos\theta'$ with
\begin{equation}
    \tan(2\theta')~:=~-\frac{2(g_{12}+2g_X)\sqrt{g_L^2+g_Y^2}\langle H\rangle^2}{\left[g_L^2+g_Y^2-\left(g_{12}+2g_X\right)^2\right]\langle H\rangle^2-4\,g_X^2\langle\Phi\rangle^2}\;.
\end{equation}
The eigenvalues are 
\begin{equation}\label{eq:mZ}
\begin{split}
        m_Z^2~=~&\frac{1}{2}(g_L^2+g_Y^2)\langle H\rangle^2-\frac{(g_{12}+2g_X)^2(g_L^2+g_Y^2)}{8g_X^2}\frac{\langle H\rangle^4}{\langle\Phi\rangle^2}+\mathcal{O}\left(\frac{\langle H\rangle^6}{\langle\Phi\rangle^4}\right),\\
        m_{Z'}^2~=~&2 g_X^2\langle\Phi\rangle^2+\frac{1}{2}(g_{12}+2g_X)^2\langle H\rangle^2+\frac{(g_{12}+2g_X)^2(g_L^2+g_Y^2)}{8g_X^2}\frac{\langle H\rangle^4}{\langle\Phi\rangle^2}+\mathcal{O}\left(\frac{\langle H\rangle^6}{\langle\Phi\rangle^4}\right).
\end{split}
\end{equation}
The tree level effective mass matrix for the scalars is given by
\begin{equation}
    \left(\begin{matrix}
        2\lambda_p H_b^2+6\lambda_\Phi\Phi_b^2 & 4 \lambda_p H_b \Phi_b \\
        4 \lambda_p H_b \Phi_b & 2\lambda_p \Phi_b^2+6\lambda_H H_b^2     \end{matrix}\right)~\oplus~\mathrm{diag}\left(2\lambda_p \Phi_b^2 + 2 \lambda_H H_b^2,2\lambda_p \Phi_b^2 + 2 \lambda_H H_b^2,2\lambda_p \Phi_b^2 + 2 \lambda_H H_b^2,2\lambda_p H_b^2 + 2 \lambda_\Phi \Phi_b^2\right)\;.
\end{equation}
While the vector boson masses are excellent approximations also after radiative corrections are taken into account, reliable expressions for the physical scalar masses (and would-be Goldstone bosons) are only obtained at the minimum of the full effective potential.

\section{Exact expression for \texorpdfstring{$\Phi_0$}{Phi0}}
Using $\langle H\rangle\ll\langle\Phi\rangle$, that is $H_b\ll\tilde{\Phi}(0):=\Phi_0$, we expand Eq.~\eqref{eq:V_eff} and the definition of $\Phi_0$ to find the exact version of \eqref{eq:Phi0_short} which reads   
\begin{equation}\label{eq:Phi0_long}
	\frac{1}{16\pi^2}\ln\left(\frac{\Phi_0^2}{\mu^2}\right)~=~-\frac{\lambda_{\Phi}+\frac{1}{16\pi^2}\left\{q_\Phi^4g_X^4 \left[3 \ln \left(2 q_\Phi^2g_X^2\right)-1\right]+4\,\lambda_p^2 \left(\ln 2\lambda_p-1\right)\right\}}{3\,q_\Phi^4g_X^4+4\,\lambda _p^2}\;.
\end{equation}	
Alternatively, we can use the $\epsilon$ expansion explained around Eq.~\eqref{eq:epsilon} in which case we redefine $\Phi_0$ as the $\mathcal{O}(\epsilon^0)$ part which reads
\begin{equation}
	\frac{1}{16\pi^2}\ln\left(\frac{\Phi_0^2}{\mu^2}\right)~=~-\frac{\lambda_{\Phi }+\frac{1}{16\pi^2}\left\{q_\Phi^4 g_X^4 \left[3 \ln \left(2q_\Phi^2 g_X^2\right)-1\right]\right\}}{ 3\, q_\Phi^4 g_X^4}\;.
\end{equation}	
This explicitly demonstrates the difference between the two expansion schemes mentioned in footnote~\ref{fn:expansion}. For our quantitative analysis, we do not use any of the expansions but perform a fully numerical minimization of the effective potential~\eqref{eq:V_eff} to compute $\langle\Phi\rangle$ and $\langle H\rangle$.

\section{Details of numerical scan}\label{app:numeric}
Here we describe the routine of our parameter scan. We randomly choose parameters in a reasonable range at the low scale, then run the model up to the high scale. 
We then impose custodially symmetric parameter relations at the high scale to subsequently run the model back down to obtain our model predictions at the low scale. 
In more detail: We randomly pick a value for the top pole mass in its $3\sigma$ range $M_t\in[170.4,174.6]\,\mathrm{GeV}$~\cite{ParticleDataGroup:2022pth}, then calculate the SM gauge and Yukawa couplings in $\overline{\text{MS}}$ at $\mu=m_t$ using the formulae of~\cite{Buttazzo:2013uya}. The couplings in the SM Higgs potential are chosen such that the SM one-loop effective potential reproduces the central values of the observed EW VEV and Higgs mass. We then randomly pick a matching scale $\tilde\mu_0\in[500,10^{6}]\,\mathrm{GeV}$ and perform a two-loop running of the SM up to $\tilde\mu_0$. We randomly chose $\left.g_X\right|_{\tilde\mu_0}\in[0,0.20]$ (larger values of $\left.g_X\right|_{\tilde\mu_0}$ would lead to a Landau pole close to or below $M_\mathrm{Pl}$). Given $\tilde\mu_0$, this fixes $\left.\lambda_\Phi\right|_{\tilde\mu_0}$. We set $\lambda_p=\left.\lambda_\Phi\right|_{\tilde\mu_0}$ and fix $\lambda_H=\left.\lambda_H^{\mathrm{SM}}\right|_{\tilde\mu_0}$, $y_t$ as well as the gauge couplings to their SM values. This fully specifies our model at $\tilde\mu_0$ and we subsequently two-loop-evolve it up to $M_\mathrm{Pl}$. At $M_\mathrm{Pl}$ we enforce $\SO{6}$ custodial symmetry by the replacement $\left.\lambda\right|_{M_\mathrm{Pl}}:=\left.\lambda_\Phi\right|_{M_\mathrm{Pl}}$, $\left.\lambda_H,\lambda_p\right|_{M_\mathrm{Pl}}\rightarrow \left.\lambda\right|_{M_\mathrm{Pl}}$. Together with $\left.g_{12}\right|_{M_\mathrm{Pl}}=0$ or $\left.g_{12}/g_X\right|_{M_\mathrm{Pl}}\in[-0.1,0.1]$ this determines a potentially good starting parameter point at the high scale. We determine $\mu_0$ and $\Phi_0$
for this parameter point by iteratively using \eqref{eq:Phi0_long} with the originally chosen $\tilde\mu_0$ as a starting point. We then numerically minimize the 1-loop effective potential at $\mu_0$ to compute the VEVs $v_\Phi$, $v_H$ and scalar masses. These are predictions of each model point and correspond to the according parameters of the SM one-loop effective potential. Hence, we match to the SM at $\mu_0$ and evolve the SM down to $\mu=m_t$. We exclude points that do not agree with the experimentally determined EW scale within $v^{\mathrm{exp}}_H\pm0.1\,\mathrm{GeV}$ (see comment in footnote~\ref{foot:G_F}),
or are in conflict with the SM values of $g_L$, $g_Y$, $g_3$ or $y_t$ within errors. 
We invert the formulae of~\cite{Buttazzo:2013uya}  to compute the top pole mass $M_t$ from our $y_t$ which is defined in $\bar{\mathrm{MS}}$.

\twocolumngrid
\bibliographystyle{utphys}
\bibliography{bib.bib}

\end{document}